\documentclass[doublespacing]{elsart}

\usepackage{amssymb}
\usepackage{epsfig}

\begin{document}

\begin{frontmatter}



\title{Analysis of the Two Dimensional Datta-Das Spin Field Effect Transistor}


\author{P. Agnihotri and S. Bandyopadhyay }
\address[label1]{Department of Electrical and Computer Engineering, 
Virginia Commonwealth 
University, Richmond, VA 23284, USA}

\begin{abstract}

An analytical expression is derived for the  conductance modulation of a ballistic {\it two-dimensional} 
Datta-Das Spin Field Effect Transistor (SPINFET) 
as a function of gate voltage. 
Using this expression, we show that the recently observed conductance modulation in a 
two-dimensional SPINFET structure does not match the theoretically expected result very well. This calls 
into question the claimed demonstration of the SPINFET 
and underscores the need for further careful investigation.

\end{abstract}

 \end{frontmatter}

\pagebreak

Even two decades after the original proposal of the Datta-Das Spin Field Effect Transistor (SPINFET) \cite{datta}, 
the exact analytical expression for the channel conductance 
of a {\it two-dimensional} device structure has remained somewhat obscure [see the Note at the end]. Although Datta and Das proposed a two-dimensional transistor 
structure in their original work \cite{datta}, the expression they derived for the channel conductance as a function of gate voltage
was based on the assumption that the carrier's wavevector 
component transverse to the direction of current flow is zero, which effectively corresponds to a one-dimensional structure.
No expression was derived for the conductance modulation in a two-dimensional structure, possibly because Datta and Das 
realized that the conductance modulation will be severely suppressed in a two-dimensional system.

Recently, a report has appeared in the literature claiming demonstration of the Datta-Das SPINFET for the first time. 
The claim is predicated on the fact that a conductance modulation was observed in a {\it two-dimensional} SPINFET structure 
as a function of gate voltage, which could be fitted exactly with the equation

\begin{equation}
\Delta G = A cos \left ( 2m^* \alpha \left [ V_G \right ] L/\hbar^2 + \phi \right ),
\end{equation}
where $\alpha \left [ V_G \right ]$ is the gate-controlled Rashba spin-orbit interaction strength 
in the two-dimensional channel, $V_G$ is the gate voltage, $m^*$ is the charge carrier's effective mass, $L$ is the source-to-drain separation 
(channel length) and $\phi$ is an arbitrary phase shift. The authors of \cite{koo} measured the expected amplitude 
$A$ and the quantity $\alpha \left [ V_G \right ]$ in their structure independently, 
and then using $\phi$ as the only fitting parameter, they could fit the experimentally observed conductance 
modulation $\Delta G$ in their structure with Equation (1). This ``fit'' (among others) was offered as proof that 
the Datta-Das transistor has been demonstrated.

Ref. \cite{koo} took Equation (1) from ref. \cite{datta}, not realizing that it applies only to a strictly 
one-dimensional channel since ref. \cite{datta} had derived it assuming that the wavevector component transverse to 
the direction of current flow is exactly zero$\footnote{The expression derived in ref. \cite{datta} did not contain 
the phase shift $\phi$, but it could arise if $\alpha \left [ V_G \right ] \neq 0$ when $V_G = 0$.}$. 
Equation (1) does {\it not}  hold for a 
two-dimensional channel since there the transverse wavevector component will not be zero. Recently, 
one of us pointed this out \cite {bandy} and derived the correct equation for a two-dimensional channel 
(of finite width) assuming that only the electron energy is conserved in ballistic transport. Subsequently, 
it was pointed out \cite{datta1} that if the width of 
the channel is semi-infinite so that periodic boundary conditions can be imposed along the width, 
then the transverse wavevector (perpendicular to the direction of current flow) is also a good quantum number
and will be conserved in ballistic transport. 
This is reminiscent of two-dimensional coherent resonant tunneling devices of semi-infinite width, 
where the transverse wavevector is conserved during tunneling \cite{capasso}. 

Conservation of the transverse wavevector greatly simplifies the equation derived in \cite{bandy}. Additionally, 
low temperature and low bias conditions cause further simplification, resulting in the following simple equation 
for the conductance modulation in a two-dimensional SPINFET:
\begin{equation}
\Delta G = B \int_0^{k_F} dk_z \left [ 1 - {{k_z^2}\over{k_F^2}} \right ] cos \left [ \Theta \left (k_F, k_z, \alpha \left [V_G \right ] 
\right ) L \right ],
\end{equation}
where $B$ is a constant and
\begin{equation}
\Theta \left (k_F, k_z, \alpha \left [V_G \right ] \right ) = - {{\left ( 2m^* \alpha \left [V_G \right ] /\hbar^2 \right ) k_F 
+ \left (m^* \right )^2 \alpha^2 \left [V_G \right ] /\hbar^4}\over{\sqrt{k_F^2 - k_z^2}}}  .
\end{equation}
Here, $k_z$ is the transverse wavevector component (along the width) and $k_F$
 is the Fermi wavevector. Derivation of Equation (2) is given in  Appendix I.

Clearly, Equation (2) has no similarity with Equation (1). Therefore, the conductance modulations in the one- and the two-dimensional 
cases are very {\it different}. Particularly, Equation (1) predicts that  the conductance modulation could reach 
100\%, whereas Equation (2) shows unambiguously that it will never reach 100\% because ensmeble averaging represented by the 
integration over the transverse wavevector component $k_z$ will dilute the modulation considerably. Only in a strictly
one-dimensional channel where Equation (1) holds, the conductance modulation can be 100\%, while in a two-dimensional channel,
it will never be 100\%.

Ref. \cite{zain} has independently derived Equation (2) for a two-dimensional SPINFET 
and found that it can be approximated as 
\begin{equation}
\Delta G \approx {{\hbar B}\over{2\sqrt{ \pi m^* \alpha \left [ V_G \right ] L}}}
cos \left [ 2m^* \alpha \left [ V_G \right ] L /\hbar^2
+ \pi/4 \right ].
\end{equation}

Equation (4) does not quite match Equation (1) either since  
the amplitude of the cosine function in Equation (4) is not constant, but 
gate-voltage dependent. Therefore, Equation (2) or Equation (4) cannot be reconciled with Equation (1).
However, ref. \cite{zain} also found that for the particular experimental 
parameters of ref. \cite{koo}, Equation (1) and Equation (2) yield similar curves for $\Delta G$ versus $V_G$ over the range of $V_G$ 
used in the experiment. This similarity is coincidental and will not be sustained over extended ranges of $V_G$. 
More importantly, we have found that if we use the values of $m^*$, $\alpha \left [V_G \right ]$, $k_F$ and $L$ reported in 
\cite{koo}, then the $\Delta G$ versus $V_G$ curve computed from the correct Equation (2) does {\it not} match the experimental 
$\Delta G$ versus $V_G$ curve reported in \cite{koo} very well. We show these two curves in Fig. 1. 
This disagreement between the correct theoretical result and the experimental observation casts doubt 
on the claimed demonstration of the Datta-Das SPINFET.

We emphasize that the above disagreement however does not establish conclusively that the Datta-Das SPINFET was not demonstrated in \cite{koo}. 
Instead, it casts doubt on the claimed demonstration and highlights the need for further investigation. 
Finally, the important question is if the observed voltage 
modulation was not due to the Datta-Das effect, what could it have been due to? Ref. \cite{koo} showed that 
the conductance modulation $\Delta G$ versus $V_G$ disappeared if the source and drain contacts were magnetized in a direction 
such that they injected and detected spins parallel to the effective magnetic field caused by the Rashba interaction. 
The modulation reappeared when the direction of magnetization was rotated by 90$^{\circ}$ so that the injected spins became 
perpendicular to the effective magnetic field. This is consistent with the Datta-Das effect which relies on precession 
of the injected spins around the effective magnetic field caused by Rashba interaction. Since precession cannot occur 
if the spins are parallel to the effective magnetic field, the Datta-Das modulation will disappear in that case. 
The precession will occur if the injected spins are perpendicular to the effective magnetic field, so that the Datta-Das effect 
is recovered when the contacts' magnetizations are rotated by 90$^{\circ}$. This observation is certainly supportive of the 
Datta-Das effect, but it could also be caused by other phenomena. One likely phenomenon is Ramsauer resonances in the channel 
\cite{cahay} which can also  give rise to a voltage modulation $\Delta G$ versus $V_G$. Ramsauer resonances are exacerbated by a magnetic 
field in the direction of current flow. In the experiment, when the contacts were magnetized in the direction  perpendicular to 
the effective magnetic field caused by the Rashba interaction, they caused a real magnetic field to appear in the channel in the direction of current flow. 
This could have induced Ramsauer resonances. When the direction of magnetization of the contacts was rotated by 90$^{\circ}$, 
the real magnetic field in the channel disappeared, which could have quenched or abated the Ramsauer resonances. Thus, the observed effect is 
also consistent with Ramsauer resonances.
Consequently, further tests are required to identifythe origin of the 
observed conductance modulation  unambiguously.
The expected oscillation periods for Ramsauer resonances and the Datta-Das effect are of course very different, but since barely one
oscillation period was observed in the experiment of ref. \cite{koo}, it is difficult to discriminate between these two effects 
from the observed modulation.

In summary, we have shown that the conductance modulation observed in ref. \cite{koo} cannot be fitted very well by the correct 
equation governing such a device, contrary to the claim of ref. \cite{koo}. Moreover, there can be alternate explanations 
for the origin of the observed conductance modulation of the device. Therefore, further careful study is required to resolve these controversies 
definitively.

\noindent {\it Note}: After the submission and acceptance of this work, we became aware of a paper [M. G. Pala, M. Governale,
J. K\"onig and U. Z\"ulicke, Europhys. Lett., {\bf 65}, 850 (2004)] which has derived an analytical expression for 
the channel conductance of a two-dimensional
SPINFET as a function of different orientations of the contacts' magnetization. That expression reduces to 
Equation (2) when the contacts are magnetized
in the +x-direction. We thank Prof. Ulrich Z\"ulicke for bringing this
to our attention.

\vfill \pagebreak

\begin{center}
{\bf Appendix I}
\end{center}

In this Appendix, we derive the expression for the channel conductance of a two-dimensional Datta-Das SPINFET as a 
function of gate voltage.

Consider the two-dimensional channel of a Spin Field Effect Transistor (SPINFET) in the x-z plane (shown in Fig. 2(a)), with current flowing 
in the x-direction. An electron's wavevector components in the channel are designated as $k_x$ and $k_z$, while the 
total wavevector  is designated as $k_t$. Note that $k_t^2 = k_x^2 + k_z^2$ as shown in Fig. 2(b).

The gate terminal induces an electric field in the y-direction which causes Rashba interaction. 
The Hamiltonian operator describing an electron in the channel is 
\begin{equation}
H = {{p_x^2 + p_z^2}\over{2m^*}} \left [ {\bf I} \right ] + \alpha \left [ V_G \right ] \left ( \sigma_z p_x - \sigma_x p_z \right ), 
\end{equation}
where the $p$-s are the momentum operators, the $\sigma$-s are the Pauli spin matrices and $\left [ {\bf I} \right ]$ is the 
2$\times$2 identity matrix. Since this Hamiltonian is invariant in both x- and z-coordinates, the wavefunctions in the 
channel are plane wave states $e^{i \left ( k_x x + k_z z \right )}$. Consequently, in the basis of these states, the Hamiltonian 
is 
\begin{equation}
H = 
\left [
\begin{array}{cc}
{{\hbar^2 k_t^2}\over{2m^*}} + \alpha \left [ V_G \right ] k_x & - \alpha \left [ V_G \right ] k_z \\
- \alpha \left [ V_G \right ] k_z & 
{{\hbar^2 k_t^2}\over{2m^*}} - \alpha \left [ V_G \right ] k_x \\
\end{array}
\right ] .
\end{equation}

Diagonalization of this Hamiltonian yields the eigenenergies and the eigenspinors in the two spin-split bands in the two-dimensional
channel:

\pagebreak

\begin{eqnarray}
E_l  =    
{{\hbar^2 k_t^2}\over{2m^*}} - \alpha \left [ V_G \right ] k_t ~ ({\tt lower~band} ); ~~~
E_u = {{\hbar^2 k_t^2}\over{2m^*}} + \alpha \left [ V_G \right ] k_t ~ ({\tt upper~band} ). \nonumber \\
\end{eqnarray}
\begin{equation}
\left [ \Psi \right ]_l  = 
\left [
\begin{array}{c}
sin \theta \\
cos \theta \\
\end{array}
\right ]~ ({\tt lower~band} ) ; ~~~
\left [ \Psi \right ]_u  = \left [
\begin{array}{c}
-cos \theta \\
sin \theta \\
\end{array}
\right ]~ ({\tt upper~band} ). 
\label{A3}
\end{equation}
where $\theta = (1/2)arctan\left ( k_z/k_x \right )$. The energy dispersion relations in the two bands (one
broken and the other solid) are plotted in Fig. 3. Note that an electron of energy $E$ has two different wavevectors in 
the two bands given by $k_t^{(1)}$ and $k_t^{(2)}$.

We will assume that the source contact of the SPINFET is polarized in the $+x$-direction and injects $+x$-polarized
spins into the channel under a source-to-drain bias. We also assume that the spin injection efficiency at the source 
is 100\%, so that only $+x$-polarized spins are injected at the complete exclusion of $-x$-polarized spins. An injected 
spin will couple into the two spin eigenstates in the channel. It is as if the 
x-polarized beam splits into two beams, each corresponding to one of the channel eigenspinors. This will yield:
\begin{eqnarray}
{{1}\over{\sqrt{2}}} 
\left [
\begin{array}{c}
1 \\
1 \\
\end{array}
\right ]
& = & C_1 \left [
\begin{array}{c}
sin \theta \\
cos \theta \\
\end{array}
\right ] + 
C_2 \left [
\begin{array}{c}
-cos  \theta \\
sin \theta \\
\end{array}
\right ], \nonumber \\
+x-polarized & & 
\label{A4}
\end{eqnarray}
where the coupling coefficients $C_1$ and $C_2$ are found by solving Equation (\ref{A4}). The result is
\begin{eqnarray}
C_1 & = & C_1 \left ( k_x, k_z \right ) = sin \left ( \theta + \pi/4 \right ) \nonumber \\
C_2 & = & C_2 \left ( k_x, k_z \right ) = -cos \left ( \theta + \pi/4 \right )
\end{eqnarray}
Note that the coupling coefficients depend on $k_x$ and $k_z$. 

At the drain end, the two beams recombine and interfere to yield the spinor of the electron impinging on the drain. 
Here, we are neglecting multiple reflection effects between the source and drain contacts in the spirit of ref. \cite{datta}.
Since the two beams have the same energy $E$ and transverse wavevector $k_z$ (these are good quantum numbers in ballistic 
transport), they must have 
different longitudinal wavevectors $k_x^{(1)}$ and $k_x^{(2)}$ since $k_t^{(1)} \neq k_t^{(2)}$.
Therefore, these 
two beams have slightly different directions of propagation in the channel. In other words, the channel behaves like a
``birefrigent'' medium where waves with anti-parallel spin polarizations travel in slightly different directions.

Hence, the spinor at the drain end will be:
\begin{eqnarray}
\left [ \Psi \right ]_{drain} & = & 
C_1 \left [
\begin{array}{c}
sin \theta \\
cos \theta \\
\end{array}
\right ]e^{i \left ( k_x^{(1)} L + k_z W \right )}
+
C_2 \left [
\begin{array}{c}
-cos \theta \\
sin \theta \\
\end{array}
\right ]e^{i \left ( k_x^{(2)} L + k_z W \right )} \nonumber \\
& = & e^{i k_z W}
\left [
\begin{array}{c}
sin \left ( \theta + \pi/4 \right ) sin \theta e^{i  k_x^{(1)} L} + 
cos \left ( \theta + \pi/4 \right ) cos \theta e^{i  k_x^{(2)} L} \\
sin \left ( \theta + \pi/4 \right ) cos \theta e^{i  k_x^{(1)} L} - 
cos \left ( \theta + \pi/4 \right ) sin \theta e^{i k_x^{(2)} L} \\
\end{array}
\right ],
\label{A6}
\end{eqnarray}
where $L$ is the channel length (distance between source and drain contacts) and $W$ is the transverse displacement of the 
electron as it traverses the channel.

Since the drain is polarized in the same orientation as the source, it transmits only +x-polarized spins, 
so that spin filtering at the drain will yield a transmission probability $|T|^2$ where $T$ is the projection  
of the impinging spinor on the eigenspinor of the drain. It is given by 
\begin{eqnarray}
T & = &  {{1}\over{\sqrt{2}}} 
\left [ 
\begin{array}{cc}
1 & 1 \\
\end{array} \right ]
\left [
\begin{array}{c}
sin \left ( \theta + \pi/4 \right ) sin \theta e^{i  k_x^{(1)} L} + 
cos \left ( \theta + \pi/4 \right ) cos \theta e^{i  k_x^{(2)} L} \\
sin \left ( \theta + \pi/4 \right ) cos \theta e^{i  k_x^{(1)} L} - 
cos \left ( \theta + \pi/4 \right ) sin \theta e^{i  k_x^{(2)} L} \\
\end{array}
\right ]e^{i k_z W} \nonumber \\
& = & e^{i k_z W} \left [ sin^2 \left ( \theta + \pi/4 \right )e^{i  k_x^{(1)} L}
+ cos^2 \left ( \theta + \pi/4 \right )e^{i k_x^{(2)} L} \right ].
\end{eqnarray}
Here, we have assumed 100\% spin filtering efficiency. 

Therefore, 
\begin{eqnarray}
|T|^2 & = &  cos^4 \left ( \theta + \pi/4 \right )	\left |1 + tan^2 \left ( \theta + \pi/4 \right ) e^{i \left [ k_x^{(1)} - k_x^{(2)}
\right ]L} \right |^2 \nonumber \\
& = & cos^4\left ( \theta + \pi/4 \right ) + sin^4\left ( \theta + \pi/4 \right )+ {{1}\over{2}}cos^2 (2 \theta ) cos (\Theta L) ,
\label{A8}
\end{eqnarray}	
where $\Theta = k_x^{(1)} - k_x^{(2)}$.

From Equation (\ref{A3}), we get that $k_t^{(1)} - k_t^{(2)} = - 2m^* \alpha \left [ V_G \right ]/\hbar^2$. 
Expressing the wavevectors in terms of their x- and z-components, we get:
\begin{equation}
\sqrt{ \left [k_x^{(1)} \right ]^2 + k_z^2} - \sqrt{ \left [k_x^{(2)} \right ]^2 + k_z^2} = - 2m^* \alpha \left [ V_G \right ]/\hbar^2,
\end{equation}
which yields
\begin{equation}
\Theta = k_x^{(1)} - k_x^{(2)} = 
{{ -2 m^* \alpha \left [ V_G \right ] /\hbar^2 k_t^{(2)}+ 2\left (m^* \right )^2 \alpha^2 \left [ V_G \right ]/\hbar^4}\over{ \left [ k_x^{(1)} + k_x^{(2)}
\right ]/2}}.
\label{A10}
\end{equation}	
		
From Equation (\ref{A3}), we also get that 
\begin{equation}
k_t^{(2)} = {{m^* \alpha \left [ V_G \right ]}\over{\hbar^2}} \pm \sqrt{ \left ( {{m^* \alpha \left [ V_G \right ]}\over{\hbar^2}} \right )^2
+ k_0^2} \approx k_0 + {{3}\over{2}} {{m^* \alpha \left [ V_G \right ] }\over{\hbar^2}} ,
\end{equation}
where $k_0 = \sqrt{2m^*E}/\hbar$. 

Now, if $\alpha \left [V_G \right ]$ is small, then $\left [ k_x^{(1)} + k_x^{(2)}
\right ]/2 \approx \sqrt{k_0^2 - k_z^2}$. Substituting these results in Equation (\ref{A10}), we get 
\begin{eqnarray}
\Theta & = &  {{ - \left ( 2m^* \alpha \left [V_G \right ] /\hbar^2 \right )k_0 - 
\left (m^* \right )^2 \alpha^2 \left [ V_G \right ]/\hbar^4}\over{\sqrt{k_0^2 - k_z^2}}} \nonumber \\
& = & {{ - \left ( 2m^* \alpha \left [V_G \right ] /\hbar^2 \right ) \sqrt{2m^*E}/\hbar 
- \left (m^* \right )^2 \alpha^2 \left [ V_G \right ]/\hbar^4}\over{\sqrt{2m^*E/\hbar^2 - k_z^2}}} .
\label{A12}
\end{eqnarray}

The current density in the channel of the SPINFET (assuming ballistic transport) is given by the Tsu-Esaki formula:
\begin{equation}
J = {{q}\over{W_y}} \int_0^{\infty} {{1}\over{h}} dE \int {{dk_z}\over{\pi}} |T|^2 \left [ f(E) - f(E + qV_{SD} ) \right ] ,
\end{equation}
where $q$ is the electronic charge, $W_y$ is the thickness of the channel (in the y-direction), $V_{SD}$ is the source-to-drain
bias voltage and $f(\eta)$ is the electron occupation probability at energy $\eta$ in the contacts. Since the contacts are at
local thermodynamic equilibrium, these probabilities are given by the Fermi-Dirac factor. 

In the linear response regime when $V_{SD} \rightarrow 0$, the above expression reduces to 
\begin{equation}
J = {{q^2V_{SD}}\over{W_y}} \int_0^{\infty} {{1}\over{h}} dE \int {{dk_z}\over{\pi}} |T|^2 \left [ - {{\partial f(E)}\over{\partial E}}
 \right ] .
\end{equation}

This yields that the channel conductance $G$ is
\begin{equation}
G = {{I_{SD}}\over{V_{SD}}} = {{J W_y W_z}\over{V_{SD}}} = {{q^2W_z}\over{\pi h }} \int_0^{\infty} dE \int dk_z |T|^2 
\left [ - {{\partial f(E)}\over{\partial E}}
 \right ] ,
\end{equation}
where $I_{SD}$ is the source-to-drain current and $W_z$ is the channel width.

Using Equations (\ref{A8}) and (\ref{A12}), we finally get that the channel conductance is 
\begin{eqnarray}
G & = &  G_0 + {{q^2W_z}\over{2 \pi h }} \int_0^{\infty} dE \int dk_z cos^2(2 \theta) cos (\Theta L) 
\left [ - {{\partial f(E)}\over{\partial E}} \right ] \nonumber \\
& = & G_0 + {{q^2W_z}\over{2 \pi h }} \int_0^{\infty} dE \int dk_z {{k_x^2}\over{k_x^2 + k_z^2}} cos (\Theta L) 
\left [ - {{\partial f(E)}\over{\partial E}} \right ] \nonumber \\
& \approx &  G_0 + {{q^2W_z}\over{2 \pi h }} \int_0^{\infty} dE \int dk_z \left [ 1 - {{\hbar^2 k_z^2}\over{2m^*E}} \right ] cos (\Theta L) 
\left [ - {{\partial f(E)}\over{\partial E}} \right ],
\end{eqnarray}
where $G_0$ is a constant independent of $\Theta$ and hence the gate voltage. It is easy to show that $G_0 = 
{{q^2W_z}\over{2 \pi h }} \int_0^{\infty} dE \int dk_z \left [ 1 + {{\hbar ^2 k_z^2}\over{2m^*E}} \right ]
\left [ - {{\partial f(E)}\over{\partial E}} \right ]$.

If the temperature is low so that $- {{\partial f(E)}\over{\partial E}} \approx \delta (E - E_F)$, then the last equation
reduces to 
\begin{eqnarray}
G & = & G_0 + {{q^2W_z}\over{2 \pi h }} \int_0^{\infty} dE \int dk_z \left [ 1 - {{\hbar^2 k_z^2}\over{2m^*E}} \right ] cos (\Theta L) 
\delta (E - E_F) \nonumber \\
& = & G_0 + {{q^2W_z}\over{2 \pi h }}  \int_0^{k_F} dk_z \left [ 1 - {{k_z^2}\over{k_F^2}} \right ] cos \left [ \Theta
\left (k_F, k_z, \alpha \left [V_G \right ] \right ) L \right ] \nonumber \\
& = & {{q^2W_z}\over{\pi h }}  \int_0^{k_F} dk_z \left [ 1 - {{k_z^2}\over{k_F^2}} \right ] F \left ( 
\alpha \left [V_G \right ], L, k_F, k_z \right ),
\end{eqnarray}
where
\begin{equation}
F \left ( 
\alpha \left [V_G \right ], L, k_F, k_z \right ) = \left \{ 
cos^2 \left [ \Theta
\left (k_F, k_z, \alpha \left [V_G \right ] \right ) L \right ]/2 + {{k_z^2}\over{k_F^2}}
sin^2 \left [ \Theta
\left (k_F, k_z, \alpha \left [V_G \right ] \right ) L \right ]/2  \right \},
\end{equation} 
and $k_F$ is the Fermi wavevector. Therefore,
\begin{equation}
\Delta G = G - G_0 = {{q^2W_z}\over{2 \pi h }}  \int_0^{k_F} dk_z \left [ 1 - {{k_z^2}\over{k_F^2}} \right ] cos \left [ \Theta
\left (k_F, k_z, \alpha \left [V_G \right ] \right ) L \right ] .
\end{equation}
The last equation is identical with Equation (2).

\vfill \pagebreak

\begin{center}
{\bf Appendix II}
\end{center}

In this appendix, we will derive an expression for $\Delta G$ assuming non-ideal spin injection and detection. Let us call 
the spin injection efficiency at the source contact $\eta_S$ and the spin filtering efficiency in the drain contact $\eta_D$.
If these efficiencies are less than 100\%, then the probability of a $+x$-polarized spin being injected by the source is 
$\left ( 1 + \eta_S \right)/2$ and the probability of it being filtered at the drain is $\left ( 1 + \eta_D \right)/2$ when both contacts
are magnetized in the +x-direction. Therefore the contribution to $\Delta G$ arising from +x-polarized injection
and +x-polarized detection is
\begin{eqnarray}
\left [ \Delta G \right ]_{+x, +x} = {{\left (1 + \eta_S \right) \left (1 + \eta_D \right)}\over{4}}{{q^2W_z}\over{2 \pi h }}  
\int_0^{k_F} dk_z \left [ 1 - {{k_z^2}\over{k_F^2}} \right ] cos \left [ \Theta
\left (k_F, k_z, \alpha \left [V_G \right ] \right ) L \right ] \nonumber \\.
\end{eqnarray}

Next consider the situation when the source injects a +x-polarized spin, but it transmits into the -x-polarized band in the drain 
because spin filtering is imperfect.

In this case, since the spin injected from the source is +x-polarized, Equation (\ref{A6}) is still valid for 
the spinor of the electron impinging on the drain. However, we now have to re-calculate the projection of the 
impinging spinor on the -x-polarized state in the drain, which will give
\begin{eqnarray}
T_{+x, -x} & = &  {{1}\over{\sqrt{2}}} 
\left [ 
\begin{array}{cc}
1 & -1 \\
\end{array} \right ]
\left [
\begin{array}{c}
sin \left ( \theta + \pi/4 \right ) sin \theta e^{i  k_x^{(1)} L} + 
cos \left ( \theta + \pi/4 \right ) cos \theta e^{i  k_x^{(2)} L} \\
sin \left ( \theta + \pi/4 \right ) cos \theta e^{i  k_x^{(1)} L} - 
cos \left ( \theta + \pi/4 \right ) sin \theta e^{i  k_x^{(2)} L} \\
\end{array}
\right ]e^{i k_z W} \nonumber \\
& = & {{1}\over{2}} e^{i k_z W} cos (2 \theta) \left [ e^{i k_x^{(2)} L} - e^{i k_x^{(1)} L} \right ].
\end{eqnarray}

This yields that 
\begin{eqnarray}
\left | T_{+x, -x} \right |^2 = {{cos^2 (2 \theta )}\over{2}} \left [ 1 - cos (\Theta L ) \right ] 
= {{1}\over{2}} \left [ 1 - {{k_z^2}\over{k_0^2}} \right ] \left \{1 - cos \left [ \Theta
\left (k_0, k_z, \alpha \left [V_G \right ] \right ) L \right ] \right \}. \nonumber \\
\end{eqnarray}
Since the probability of injecting a +x-polarized spin  at the source contact is $\left (1 + \eta_S \right)/2$
and the probability of its transmitting into the -x-polarized band at the drain contact is
$ \left (1 - \eta_D \right)/2$, the corresponding contribution to $\Delta G$ will be  
\begin{eqnarray}
\left [ \Delta G \right ]_{+x, -x} = {{\left (1 + \eta_S \right) \left (1 - \eta_D \right)}\over{4}}{{q^2W_z}\over{2 \pi h }}  
\int_0^{k_F} dk_z \left [ 1 - {{k_z^2}\over{k_F^2}} \right ] \{ 1 - cos \left [ \Theta
\left (k_F, k_z, \alpha \left [V_G \right ] \right ) L \right ] \}. \nonumber \\
\end{eqnarray}

Now, consider the situation when the source injects a -x-polarized spin, but it transmits into the +x-polarized band in the drain.

In this case, the spin injected from the source contact is -x-polarized and we will have to recalculate 
the spinor of the electron impinging on the drain. Equation (\ref{A4}) will now be replaced by
\begin{equation}
{{1}\over{\sqrt{2}}} 
\left [
\begin{array}{c}
1 \\
-1 \\
\end{array}
\right ]
= C_1^{\prime} \left [
\begin{array}{c}
sin \theta \\
cos \theta \\
\end{array}
\right ] + 
C_2^{\prime} \left [
\begin{array}{c}
-cos  \theta \\
sin \theta \\
\end{array}
\right ],
\end{equation}		

which yields
\begin{eqnarray}
C_1^{\prime} & = & sin \left ( \theta - \pi/4 \right ) \nonumber \\
C_2^{\prime} & = & -cos \left ( \theta - \pi/4 \right ) .
\end{eqnarray}

Therefore, the spinor of the electron impinging on the drain is
\begin{eqnarray}
\left [ \Psi \right ]_{drain} & = & 
C_1^{\prime} \left [
\begin{array}{c}
sin \theta \\
cos \theta \\
\end{array}
\right ]e^{i \left ( k_x^{(1)} L + k_z W \right )}
+
C_2^{\prime} \left [
\begin{array}{c}
-cos \theta \\
sin \theta \\
\end{array}
\right ]e^{i \left ( k_x^{(2)} L + k_z W \right )} \nonumber \\
& = & e^{i k_z W}
\left [
\begin{array}{c}
sin \left ( \theta - \pi/4 \right ) sin \theta e^{i  k_x^{(1)} L} + 
cos \left ( \theta - \pi/4 \right ) cos \theta e^{i  k_x^{(2)} L} \\
sin \left ( \theta - \pi/4 \right ) cos \theta e^{i  k_x^{(1)} L} - 
cos \left ( \theta - \pi/4 \right ) sin \theta e^{i  k_x^{(2)} L} \\
\end{array}
\right ], \nonumber \\
\label{S8}
\end{eqnarray}

The projection of this spinor on the +x-polarized state in the drain gives
\begin{eqnarray}
T_{-x, +x} & = &  {{1}\over{\sqrt{2}}} 
\left [ 
\begin{array}{cc}
1 & 1 \\
\end{array} \right ]
\left [
\begin{array}{c}
sin \left ( \theta - \pi/4 \right ) sin \theta e^{i  k_x^{(1)} L} + 
cos \left ( \theta - \pi/4 \right ) cos \theta e^{i  k_x^{(2)} L} \\
sin \left ( \theta - \pi/4 \right ) cos \theta e^{i  k_x^{(1)} L} - 
cos \left ( \theta - \pi/4 \right ) sin \theta e^{i  k_x^{(2)} L} \\
\end{array}
\right ]e^{i k_z W} \nonumber \\
& = & {{1}\over{2}} e^{i k_z W} cos (2 \theta) \left [ e^{i  k_x^{(2)} L} - e^{i k_x^{(1)} L} \right ].
\end{eqnarray}

Therefore, once again, 
\begin{eqnarray}
\left | T_{+x, -x} \right |^2 & = & {{cos^2 (2 \theta )}\over{2}} \left [ 1 - cos (\Theta L ) \right ] \nonumber \\
& = & {{1}\over{2}} \left [ 1 - {{k_z^2}\over{k_0^2}} \right ] \left \{1 - cos \left [ \Theta
\left (k_0, k_z, \alpha \left [V_G \right ] \right ) L \right ] \right \}.
\end{eqnarray}

Since the probability of injecting a -x-polarized spin at the source is $\left (1 - \eta_S \right)/2$
and the probability of its transmitting into the +x-polarized band at the drain is $\left (1 + \eta_D \right)/2$, the corresponding 
contribution to $\Delta G$ is
\begin{eqnarray}
\left [ \Delta G \right ]_{-x, +x} = {{\left (1 - \eta_S \right) \left (1 + \eta_D \right)}\over{4}}{{q^2W_z}\over{2 \pi h }}  
\int_0^{k_F} dk_z \left [ 1 - {{k_z^2}\over{k_F^2}} \right ] \{ 1 - cos \left [ \Theta
\left (k_F, k_z, \alpha \left [V_G \right ] \right ) L \right ] \}. \nonumber \\
\end{eqnarray}

Finally, consider the situation when a -x-polarized electron is injected at the source and transmits into the -x-polarized band of the 
drain 
contact.

In this case, Equation (\ref{S8}) will describe the spinor of the electron impinging on the drain and the projection of this spinor 
on the -x-polarized state in the drain will give 
\begin{eqnarray}
T_{-x, -x} & = &  {{1}\over{\sqrt{2}}} 
\left [ 
\begin{array}{cc}
1 & -1 \\
\end{array} \right ]
\left [
\begin{array}{c}
sin \left ( \theta - \pi/4 \right ) sin \theta e^{i  k_x^{(1)} L} + 
cos \left ( \theta - \pi/4 \right ) cos \theta e^{i  k_x^{(2)} L} \\
sin \left ( \theta - \pi/4 \right ) cos \theta e^{i  k_x^{(1)} L} - 
cos \left ( \theta - \pi/4 \right ) sin \theta e^{i  k_x^{(2)} L} \\
\end{array}
\right ]e^{i k_z W} \nonumber \\
& = &  e^{i k_z W}  \left \{ sin^2 \left ( \theta - \pi/4 \right )e^{i  k_x^{(1)} L} + 
cos^2 \left ( \theta - \pi/4 \right )e^{i  k_x^{(2)} L} \right \}.
\end{eqnarray}

Consequently, the contribution to $\Delta G$ will be 

\begin{eqnarray}
\left [ \Delta G \right ]_{-x, -x} = {{\left (1 - \eta_S \right) \left (1 - \eta_D \right)}\over{4}}{{q^2W_z}\over{2 \pi h }}  
\int_0^{k_F} dk_z \left [ 1 - {{k_z^2}\over{k_F^2}} \right ]  cos \left [ \Theta
\left (k_F, k_z, \alpha \left [V_G \right ] \right ) L \right ] \nonumber \\.
\end{eqnarray}

Using the Principle of Superposition, the total gate voltage dependent conductance modulation will be 
\begin{eqnarray}
\Delta G & = & \left [{{\left (1 + \eta_S \right) \left (1 + \eta_D \right)}\over{4}} -
{{\left (1 + \eta_S \right) \left (1 - \eta_D \right)}\over{4}} -
{{\left (1 - \eta_S \right) \left (1 + \eta_D \right)}\over{4}} 
+ {{\left (1 - \eta_S \right) \left (1 - \eta_D \right)}\over{4}} \right ] \nonumber \\
&& \times  {{q^2W_z}\over{2 \pi h }}  
\int_0^{k_F} dk_z \left [ 1 - {{k_z^2}\over{k_F^2}} \right ]  cos \left [ \Theta
\left (k_F, k_z, \alpha \left [V_G \right ] \right ) L \right ] \nonumber \\
& = & \eta_S \eta_D {{q^2W_z}\over{2 \pi h }}  
\int_0^{k_F} dk_z \left [ 1 - {{k_z^2}\over{k_F^2}} \right ]  cos \left [ \Theta
\left (k_F, k_z, \alpha \left [V_G \right ] \right ) L \right ].
\end{eqnarray}

Therefore, non-ideal spin injection and filtering reduces the amplitude of any non-local voltage modulation by the factor $\eta_S \eta_D$.

\vfill \pagebreak


\begin{thebibliography}{10}


\bibitem{datta}
S. Datta and B. Das, Appl. Phys. Lett., {\bf 56}, 665 (1990).

\bibitem{koo}
H. C. Koo, J. H. Kwon, J. Eom, J. Chang, S. H. Han and M. Johnson, Science, {\bf 325}, 1515 (2009).


\bibitem{bandy}

S. Bandyopadhyay, arXiv:cond-mat/0911.0210.

\bibitem{datta1}
S. Datta, private communication.

\bibitem{capasso}
F. Capasso, S. Sen, F. Beltram and A. Y. Cho, in Physics of Quantum Electron Devices, 
Springer Series in Electronics and Photonics, Vol. 28, Ed. F. Capasso, (Springer-Verlag, Berlin-Heidelberg, 1990), Chapter 7.

\bibitem{zain}

A. N. M. Zainuddin,  S. Hong, L. Siddiqui and S. Datta, arXiv:cond-mat/1001:1523.

\bibitem{cahay}
M. Cahay and S. Bandyopadhyay, Phys. Rev. B., {\bf 68}, 115316 (2003).


\end{thebibliography}

{\vfill \pagebreak}

\begin{figure}[h]
\epsfxsize=6.3in
\epsfysize=4.3in
\centerline{\epsffile{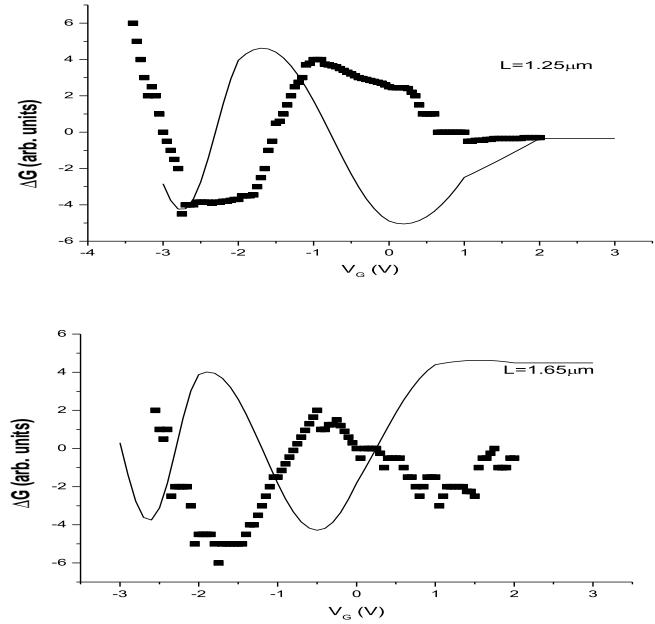}}
\caption[Fig. 1]{Plots of channel conductance modulation $\Delta G$ of a two-dimensional SPINFET versus gate voltage $V_G$. The 
solid lines are theoretical results calculated from Equation (2) where we have used the values of $m^*$, $k_F$ and 
$\alpha \left [ V_G \right ]$ reported in reference \cite{koo} and the points are (approximated) experimental results reported in \cite{koo}.
The amplitudes of the theoretical plots are adjusted to match the experimental results as closely as possible. Note that neither the 
periods, nor the phases of the experimental plots agree very well with the theoretical plots. The plots are for two different channel 
lengths of $L$ = 1.65 $\mu$m and 1.25 $\mu$m used in the experiments of ref. \cite{koo}. The experiments were carried out at low temperatures and biasesThe spin splitting energy in the dot specified in Fig. 2 as a function of 
Rashba interaction strength.}
\end{figure}

\begin{figure}[h]
\epsfxsize=6.3in
\epsfysize=4.3in
\centerline{\epsffile{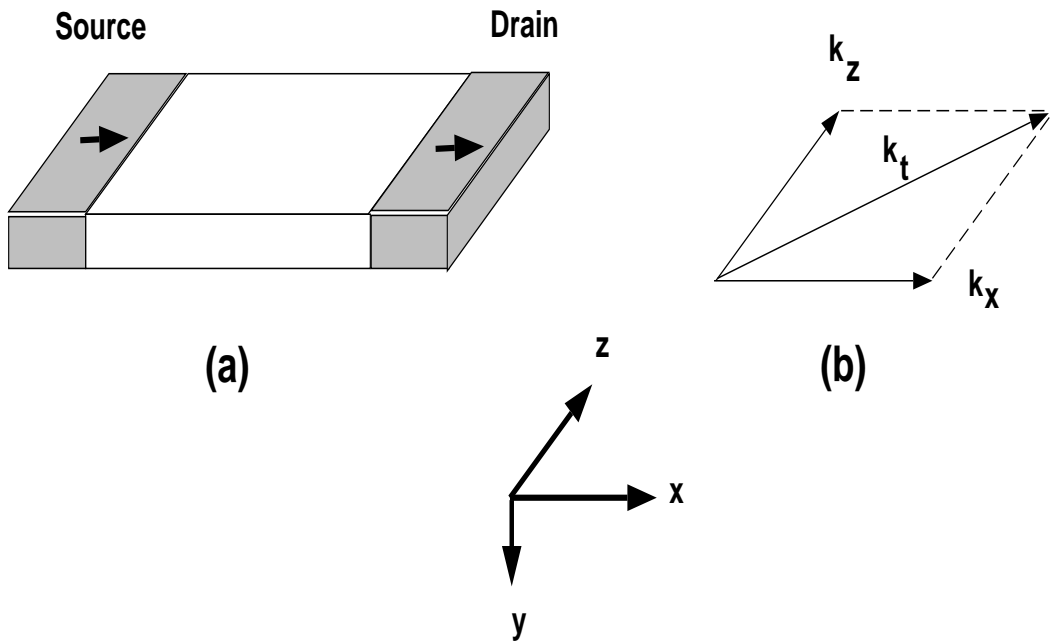}}
\caption[Fig. 2]{ (a) A two-dimensional SPINFET channel, and (b) the wavevector components in the plane of the channel. }
\end{figure}

\begin{figure}[h]
\epsfxsize=6.3in
\epsfysize=4.3in
\centerline{\epsffile{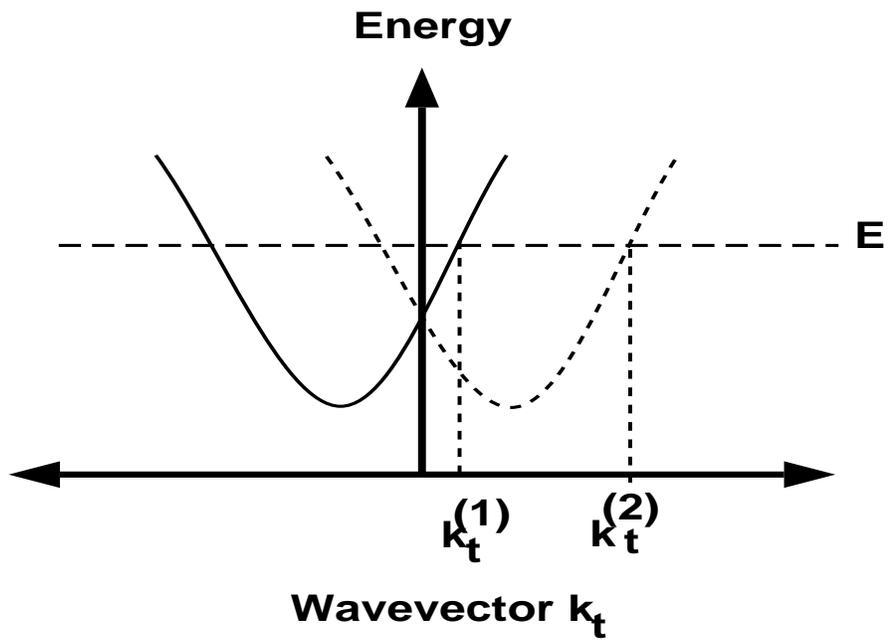}}
\caption[Fig. 3]{ Schematic representation of the dispersion relations in the two spin split bands, under the influence of the 
gate voltage inducing Rashba interaction in the channel.}
\end{figure}

\end{document}